\documentclass[letterpaper]{article} % DO NOT CHANGE THIS
\usepackage{aaai25}  % DO NOT CHANGE THIS
\usepackage{times}  % DO NOT CHANGE THIS
\usepackage{helvet}  % DO NOT CHANGE THIS
\usepackage{courier}  % DO NOT CHANGE THIS
\usepackage[hyphens]{url}  % DO NOT CHANGE THIS
\usepackage{graphicx} % DO NOT CHANGE THIS
\usepackage{natbib}  % DO NOT CHANGE THIS AND DO NOT ADD ANY OPTIONS TO IT
\usepackage{caption} % DO NOT CHANGE THIS AND DO NOT ADD ANY OPTIONS TO IT
\usepackage{algorithm}
\usepackage{algorithmic}

\usepackage{subcaption}
\usepackage{multirow}
\usepackage{amsmath}
\usepackage{pdfpages}
\usepackage{comment}
\usepackage{lipsum}
\usepackage{etoolbox}
\usepackage{array}
\usepackage{adjustbox}
\usepackage{soul}
\usepackage[utf8]{inputenc}
\usepackage[T1]{fontenc}
\usepackage{xcolor}
\usepackage{lineno}
\usepackage{newfloat}
\usepackage{listings}
\DeclareCaptionStyle{ruled}{labelfont=normalfont,labelsep=colon,strut=off} % DO NOT CHANGE THIS
\floatstyle{ruled}
\newfloat{listing}{tb}{lst}{}
\floatname{listing}{Listing}
\title{Partisan Fact-Checkers’ Warnings Can Effectively Correct Individuals’ Misbeliefs About Political Misinformation}% \xiong{add labels or warnings after fact-checkers, e.g., Fact-Checker Warnings?}}
\author{
    Written by AAAI Press Staff\textsuperscript{\rm 1}\thanks{With help from the AAAI Publications Committee.}\\
    AAAI Style Contributions by Pater Patel Schneider,
    Sunil Issar,\\
    J. Scott Penberthy,
    George Ferguson,
    Hans Guesgen,
    Francisco Cruz\equalcontrib,
    Marc Pujol-Gonzalez\equalcontrib
}
\affiliations{
    \textsuperscript{\rm 1}The University of Mississippi, USA\\
    \textsuperscript{\rm 2}The Pennsylvania State University, USA\\
    slee63@olemiss.edu, hxs378@psu.edu, axx29@psu.edu, dongwon@psu.edu
}

\title{Partisan Fact-Checkers’ Warnings Can Effectively Correct Individuals’ Misbeliefs About Political Misinformation}
\author {
    Sian Lee\textsuperscript{\rm 1},
    Haeseung Seo\textsuperscript{\rm 2},
    Aiping Xiong\textsuperscript{\rm 2},
    Dongwon Lee\textsuperscript{\rm 2}
}
\usepackage{bibentry}
\begin{document}

\maketitle

\begin{abstract}
Political misinformation, particularly harmful when it aligns with individuals' preexisting beliefs and political ideologies, has become widespread on social media platforms. In response, platforms like Facebook and X introduced warning messages leveraging fact-checking results from third-party fact-checkers to alert users against false content. However, concerns persist about the effectiveness of these fact-checks, especially when fact-checkers are perceived as politically biased. To address these concerns,
this study presents findings from an online human-subject experiment (N=216) investigating how the political stances of fact-checkers influence their effectiveness in correcting misbeliefs about political misinformation.
Our findings demonstrate that partisan fact-checkers can decrease the perceived accuracy of political misinformation and correct misbeliefs without triggering backfire effects. This correction is even more pronounced when the misinformation aligns with individuals' political ideologies. Notably, while previous research suggests that fact-checking warnings are less effective for conservatives than liberals, our results suggest that explicitly labeled partisan fact-checkers, positioned as political counterparts to conservatives, are particularly effective in reducing conservatives' misbeliefs toward pro-liberal misinformation.
\end{abstract}
\section{Introduction}

The widespread use of social media for news consumption has played a significant role in the dissemination of misinformation and fake news~\cite{allcott2017social,lazer2018science}. During critical events such as the COVID-19 pandemic, misinformation has contributed to harmful outcomes, including vaccine hesitancy and the adoption of unproven treatments~\cite{caceres2022impact}. It can also shape public opinion on key societal issues and distort individuals’ perceptions of facts and beliefs~\cite{meng2023impact}. Understanding the factors that make people susceptible to misinformation, as well as strategies to correct misbeliefs, is essential for mitigating its impact and preventing future harm.

Misinformation becomes particularly pervasive when it involves political topics—often referred to as political misinformation~\cite{jerit2020political}. This type of misinformation is especially persistent because it aligns with people's preexisting beliefs and political ideologies~\cite{bode2015related,garrett2013undermining,jerit2012partisan,taber2006motivated}, making it challenging to correct the associated misbeliefs~\cite{jerit2020political,morris2020fake,prike2023effective,walter2020fact}.

In response to the spread of misinformation on social media, researchers and platforms like Facebook and Twitter (now `X') have explored various strategies to counter misinformation, including the use of warning messages~\cite{bode2018see, clayton2020real, pennycook2018prior, seo2019trust}. To achieve this, platforms often collaborate with independent fact-checkers certified by the International Fact-Checking Network (IFCN)\footnote{https://ifcncodeofprinciples.poynter.org/signatories}. For instance, Meta\footnote{https://www.facebook.com/formedia/blog/third-party-fact-checking-how-it-works. As of April 7, 2025, Meta updated its policy to use third-party fact-checkers only outside the U.S., while piloting a community-based system (Community Notes) within the U.S.} partners with IFCN-certified fact-checkers to add warning messages that alert users to misinformation.

Despite the IFCN’s emphasis on non-partisanship, research in academia~\cite{marietta2015fact,mena2019principles,soprano2024cognitive} and analyses by media outlets such as \textit{AllSides}\footnote{https://www.allsides.com/media-bias/fact-check-bias-chart} and \textit{Ad Fontes Media}\footnote{https://adfontesmedia.com/interactive-media-bias-chart/} suggest that some fact-checkers may exhibit political biases. \textit{AllSides}, for example, evaluates fact-checker bias and incorporates user feedback to update its bias chart, which reflects users’ perceptions of political bias in fact-checkers and potentially influences their views of the fact-checkers’ credibility.~\cite{van2020you}. In such cases, individuals may place greater trust in fact-checkers whose political stances align with their own, rather than those with opposing views~\cite{van2020you}, potentially impacting both the perceived credibility of the fact-checkers and the effectiveness of their corrections~\cite{guillory2013correcting,prike2023effective,seo2022if}. %\textcolor{blue}{
For instance, Meta recently announced the removal of third-party fact-checkers on Facebook and Instagram, citing concerns about political bias of the fact-checkers and its impact on trust. Meta CEO Mark Zuckerberg acknowledged that while this move aims to reduce perceived bias, it could potentially allow harmful content to appear~\cite{suciu2025meta}.
Furthermore, prior research indicates that the effectiveness of corrections to misinformation can vary by political ideology, with conservatives possibly less influenced by fact-checking messages than liberals~\cite{morris2020fake}. %}
However, the influence of political stance congruency between fact-checkers and the individuals consuming their messages, particularly regarding the effectiveness of corrections, remains underexplored. Addressing this gap, and given the limited prior research on how a fact-checker’s political stance impacts the efficacy of fact-checking messages in polarized contexts, this study investigates the following three research questions (\textbf{RQs}):

\begin{itemize}
    \item \textbf{RQ1} [Correction Effectiveness by Fact-Checker Congruence]: Do participants reduce more misbeliefs in political misinformation when the warning message comes from a fact-checker with a politically congruent stance, compared to one with a politically incongruent stance? 
    \item \textbf{RQ2} [Correction Effectiveness by Misinformation Congruence]: Does the effectiveness of the correction differ between congruent and incongruent political misinformation?
    \item \textbf{RQ3} [Correction Effectiveness by Participant Ideology]: Does the effectiveness of fact-checking corrections vary between liberal and conservative participants?
\end{itemize}

We conducted an online experiment to examine whether fact-checking messages from partisan fact-checkers can correct misbeliefs (\textbf{RQ1}) without triggering a backfire effect, a phenomenon where corrections unintentionally strengthen misbeliefs instead of reducing them~\cite{nyhan2010corrections}. We also explored how the political stance of misinformation influences perceived accuracy and interacts with participants’ political ideologies, affecting the effectiveness of fact-checking messages (\textbf{RQ2}). Lastly, we analyzed differences between liberals and conservatives in the effectiveness of corrections (\textbf{RQ3}). 

Our  results showed that fact-checkers, regardless of their political bias, effectively reduced misbeliefs about misinformation (\textbf{RQ1}) without causing backfire effects. Furthermore, fact-checking was more effective for congruent misinformation than for incongruent misinformation (\textbf{RQ2}). For conservatives, corrections from politically incongruent fact-checkers were particularly effective in reducing misbeliefs (\textbf{RQ3}), although this effect was not observed for liberals.
\section{Related Work}

\subsection{Misinformation on Social Media: The Impact of Fact-Checking Warning Messages}

Research has demonstrated that warning messages on social media can effectively reduce belief in misinformation~\cite{clayton2020real, lu2022effects, martel2023misinformation, pennycook2018prior, seo2019trust, yaqub2020effects}. For instance, Facebook’s warning labels—such as “Disputed” and “False Information – Checked by independent fact-checkers”—have been shown to decrease both sharing intentions and the perceived accuracy of misinformation~\cite{martel2023misinformation,pennycook2018prior,seo2019trust}. \citeauthor{bode2015related} (\citeyear{bode2015related}) investigated the effect of Facebook’s “Related Articles” feature, which presents additional links alongside posts, on the perception of misinformation. Their findings suggest that when these related articles include corrective information, users’ belief in misinformation is significantly reduced. Building on these insights, this study embeds both warning labels and related-article-style correction messages into its experimental stimuli (see Figure~\ref{fig:fbpost}) to assess their effectiveness in reducing belief in misinformation.

\subsection{Political Misinformation: Political Stance Congruency and Its Correction}
Research consistently shows that individuals are more susceptible to misinformation that aligns with their political beliefs, highlighting the impact of political-stance congruency on misinformation~\cite{frenda2013false,gao2018label,xiong2022effects}. This congruency often makes it difficult for individuals to update their beliefs, even when faced with corrections ~\cite{bode2015related,garrett2013undermining,jerit2012partisan,taber2006motivated}. However, studies examining the effectiveness of corrections in addressing political misinformation have yielded mixed results.~\cite{jerit2020political,prike2023effective,walter2020fact}.

Some studies have found that corrections can effectively reduce belief in both politically congruent and incongruent misinformation, suggesting their potential to mitigate misbeliefs regardless of partisan alignment~\cite{swire2017processing,swire2020they}. For example, \citeauthor{swire2020they} (\citeyear{swire2020they})  conducted an experiment with 1,501 U.S. residents and found that supporters of Trump or Sanders were more likely than non-supporters to believe statements made by their respective politicians, regardless of whether the statements were factual or misinformation. However, correction messages effectively reduced belief in misinformation from both politicians, regardless of the participants' alignment. Similarly, \citeauthor{nyhan2020taking} (\citeyear{nyhan2020taking}) examined responses to fact-checks of Donald Trump's claims from his 2016 convention speech and a general election debate. They found that Trump supporters believed his claims more than Clinton supporters, but fact-checking reduced misinformation beliefs for both groups, with Trump supporters adjusting their beliefs less. Additionally, \citeauthor{Hameleers2020} (\citeyear{Hameleers2020}) showed that fact-checking messages can significantly reduce agreement with politically congruent misinformation and help mitigate political polarization.

Conversely, other studies suggest that the impact of corrections on political misinformation is limited and varies based on individuals' political ideologies~\cite{jennings2023asymmetric,morris2020fake,yaqub2020effects}. For instance, \citeauthor{morris2020fake} (\citeyear{morris2020fake}) conducted an experiment with 1,284 participants using fact-checking messages from a non-partisan source on news stories critical of either Democrats or Republicans. They found that participants' likelihood of recognizing the truth was mainly influenced by the consistency of the information with their preexisting partisan and ideological beliefs, with conservatives being less persuaded by fact-checking messages than liberals. The researchers suggested that this difference may stem from varying levels of trust in experts and institutions that present themselves as nonpartisan, particularly in the context of a highly polarized political environment.~\cite{morris2020fake}.

Additionally, \citeauthor{nyhan2010corrections} (\citeyear{nyhan2010corrections}) found that fact-checking corrections on political misinformation, such as the claim that Iraq had Weapons of Mass Destruction (WMD), were not only ineffective but sometimes backfired, reinforcing misbeliefs among some conservatives who were strongly aligned with the misinformation. However, more recent studies suggest that fact-checking is unlikely to trigger a backfire effect, even on highly polarized issues~\cite{ecker2020can, prike2023examining, wood2019elusive}. 
\citeauthor{swire2020searching} (\citeyear{swire2020searching}) present a comprehensive review of the backfire effect literature, concluding that it is not a robust empirical phenomenon. Their findings reassure practitioners that fact-checking rarely leads to increased belief in misinformation at the group level. The authors further emphasize the importance of employing rigorous methodologies and delivering clear corrective messaging to maximize the effectiveness of fact-checking efforts.

\subsection{Source Credibility and Media Bias}

Research shows that source credibility significantly impacts the effectiveness of corrections~\cite{guillory2013correcting, martel2023misinformation, prike2023effective, seo2022if, vraga2018not}. For instance, \citeauthor{vraga2018not} (\citeyear{vraga2018not}) found that corrections paired with credible sources on platforms like Facebook and Twitter (now `X') effectively reduce misperceptions about misinformation. Similarly, \citeauthor{seo2022if} (\citeyear{seo2022if}) demonstrated that source credibility influences participants' acceptance of corrections on COVID-19 misinformation. These findings underscore the importance of how fact-checkers are perceived when delivering fact-checking messages.

Media, including fact-checkers, can exhibit political bias in various ways~\cite{marietta2015fact,mena2019principles,soprano2024cognitive}. For instance, they may show coverage bias by predominantly reporting negative news about a specific party or ideology, such as frequently labeling statements from certain politicians as false~\cite{d2000media,eberl2017one}. They may also exhibit agenda-setting bias by focusing on particular political figures and topics that align with their favored policies~\cite{eberl2017one,brandenburg2006party,hofstetter1978bias,groeling2013media}. Despite these biases, media and fact-checkers can still provide accurate information. 

\citeauthor{jia2024journalistic} (\citeyear{jia2024journalistic}) further highlight the importance of human expertise in fact-checking, showing that fact-checking labels created by professional fact-checkers or journalists are perceived as more effective than those generated by algorithms or users. This highlights the lasting importance of human fact-checkers, even amidst the rapid advancement of automated fact-checking systems. Their finding also emphasizes the critical role of human judgment in the fact-checking process, despite the potential for inherent biases in their evaluations.

\citeauthor{swire2017processing} (\citeyear{swire2017processing}) examined the impact of fact-checking sources by categorizing them into three conditions: `according to Democrats,' `according to Republicans,' and `according to a non-partisan fact-checking website,' focusing on misinformation attributed to Donald Trump. They found that the source's political stance had minimal impact on the effectiveness of fact-checking messages, with significance observed only among Republican non-supporters of Trump, but not for Republican supporters of Trump or Democrats. Post-hoc analysis showed that corrections from Republicans further reduced the perceived accuracy of misinformation for these participants compared to corrections from Democrats or non-partisan sources. This led the authors to reject their hypothesis that corrections from `unlikely sources' (e.g., a Republican correcting misinformation from another Republican) would be more effective. This finding contrasts with \citeauthor{berinsky2017rumors} (\citeyear{berinsky2017rumors}), who found that countering political rumors with corrections from unlikely sources enhances individuals' readiness to dismiss such rumors, making unlikely sources more effective than likely ones, regardless of political ideology. 

However, their experimental setting accounted for not only the political stance of the corrections (i.e., whether they came from Republicans or Democrats), which influenced the perceived trustworthiness of the source across different political ideologies, but also the source’s expertise (e.g., third-party fact-checking websites versus politicians with vested interests in Donald Trump’s statements) as factors affecting the effectiveness of corrections~\cite{yaqub2020effects}. Consequently, the study conflated two aspects of credibility—trustworthiness and expertise—within a single independent variable of the fact-checking message's source, complicating the assessment of the impact of fact-checking messages~\cite{mcginnies1980better}. Additionally, their study was limited to statements from Donald Trump and did not consider statements from other political figures. 

In our study, we standardized the source of fact-checking messages to a fact-checker, varying only the political stance of the fact-checker (leaning either liberal or conservative). 
In the context of fact-checking, the perceived political alignment between fact-checkers and fact-checked claims can significantly shape how the corrections are interpreted (also known as the messenger effect; McGinnies and Ward 1980; Petty and Cacioppo 1986), particularly when the fact-checker's political ideology aligns with the recipient's ideological stance. This study investigates this dynamic in the context of corrections made by partisan fact-checkers on social media, a topic that remains underexplored in the literature.

\section{Present Study}

Political misinformation is particularly persistent, and previous research shows mixed results regarding the effectiveness of fact-checking messages in correcting misbeliefs associated with it. The effectiveness of the corrections often depends on the perceived credibility of the source, such as fact-checkers, who may be viewed as politically biased. Since perceptions of fact-checkers can vary based on individuals' political ideologies, the effectiveness of corrections from fact-checkers can differ among people, even when the correction messages are identical.

While studies show that fact-checkers frequently review the same misinformation and usually agree on their verdicts~\cite{amazeen2015revisiting,amazeen2016checking,lee2023fact}, it remains unclear which fact-checker is most effective when multiple sources debunk the same misinformation on social media. This study examines how the partisan identity of fact-checkers influences the effectiveness of misinformation corrections. While prior research has established the general efficacy of fact-checking \cite{amazeen2015revisiting, clayton2020real, martel2023misinformation, nyhan2010corrections, pennycook2018prior}, this work uniquely contributes by investigating the underexplored role of a fact-checker’s perceived political stance.

\section{Method}
\subsection{Materials} 
For our main study, we selected six real and six fake headlines related to U.S. politics, with each set containing three pro-liberal and three pro-conservative headlines, ensuring an even split of viewpoints within both the real and fake categories. 
All headlines were presented in a standardized format: text-based, standalone (without accompanying images), concise (one to three sentences), and free of grammatical errors or expressive punctuation (e.g., exclamation marks). This standardization ensured that participants could not use superficial cues to distinguish real headlines from fake ones.
Furthermore, we intentionally chose items that were not too recent, published on or before June 2023, to reduce the likelihood that people would clearly remember the content.

To achieve this, the first author initially selected 12 real and 12 fake headlines, labeling each for political bias (pro-liberal or pro-conservative). Other authors then independently labeled the political bias of all 24 headlines. If there was agreement among all authors, the headline was retained; if not, the first author replaced the disputed headline with a new one. This iterative process continued until 24 consistently labeled headlines were confirmed. 
We then conducted a pretest~\cite{pennycook2020implied,xiong2022effects} on the crowdsourcing platform Prolific to identify 12 headlines that met two criteria: 1) validation of the accuracy of our political bias labels (pro-liberal or pro-conservative), and 2) balanced strength of political leanings between pro-liberal and pro-conservative headlines across both real and fake news categories. Based on the pretest results, we selected 6 real headlines and 6 fake headlines, evenly divided between pro-liberal and pro-conservative, for use in our main study. The supplementary material\footnote{Supplementary material is available at tiny.cc/leeetal25supp} provides further details on how the pretest was conducted.

\begin{figure}[h!]%[h!]
  \centering
  \includegraphics[width=0.47\textwidth]{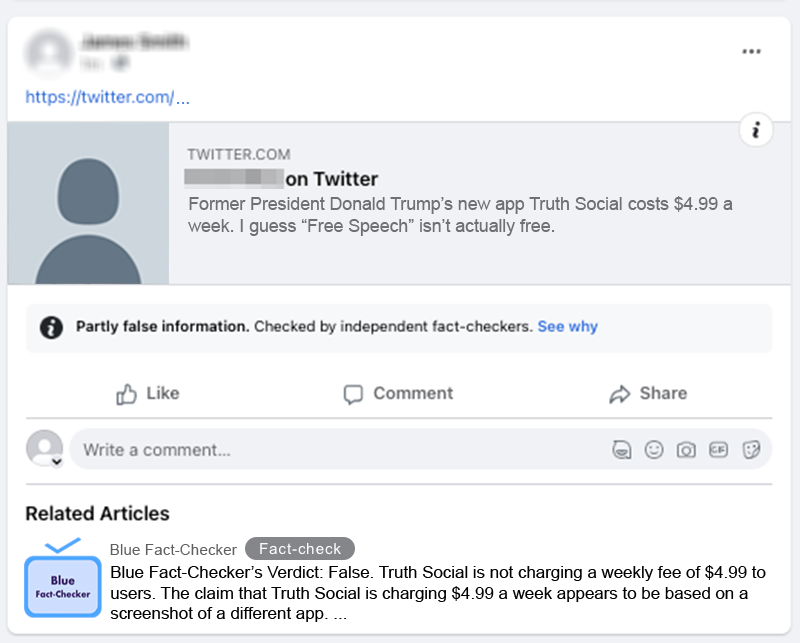}
  \caption{Example of the experimental stimulus under the \textit{Blue Fact-Checker} condition with a fake headline. The \textit{Red Fact-Checker} condition is identical, except it includes a red fact-checker icon and the label ``Red Fact-Checker.'' In the \textit{No Fact-Checker} condition, no warning tags or related articles are shown and only the shared social media link with the headline is presented. The headline content remains the same across all three fact-checker conditions. For real headline stimuli, the format is the same as the \textit{No Fact-Checker} condition, but displays a real headline instead.}
  \label{fig:fbpost}
\end{figure}

Real headlines were sourced from reputable outlets such as \textit{CNN} and \textit{Fox News}, ensuring that none had been previously debunked by fact-checkers. Fake headlines were obtained from fact-checking websites like \textit{Snopes}, \textit{PolitiFact}, and \textit{CheckYourFact.com}, selecting only those that had been debunked by both left-leaning and right- (or center-) leaning fact-checkers, according to the AllSides Fact Check Bias Chart version 3.0\footnote{https://www.allsides.com/media-bias/fact-check-bias-chart}. This approach was intended to enhance the ecological validity of our study by using the same fact-checking warning message from either a left-leaning (Blue Fact-Checker) or right-leaning (Red Fact-Checker) fact-checker. The supplementary material provides further details on the selection process for the fake headlines. To further enhance ecological validity, each headline was incorporated into a Facebook post format, depicting a user sharing a link from another social media post containing that headline (see Figure~\ref{fig:fbpost}).

\begin{table}[]
\centering
\resizebox{\columnwidth}{!}{%
\begin{tabular}{|r|l|rr|}
\hline
\multicolumn{1}{|c|}{\multirow{2}{*}{Item}} &
  \multicolumn{1}{c|}{\multirow{2}{*}{Options}} &
  \multicolumn{2}{c|}{Participant Political Stance} \\ \cline{3-4} 
\multicolumn{1}{|c|}{} &
  \multicolumn{1}{c|}{} &
  \multicolumn{1}{c|}{\begin{tabular}[c]{@{}c@{}}Liberals\\ (N=110)\end{tabular}} &
  \multicolumn{1}{c|}{\begin{tabular}[c]{@{}c@{}}Conservatives\\ (N=106)\end{tabular}} \\ \hline
\multirow{4}{*}{Gender} & Female                           & \multicolumn{1}{r|}{60.0\%} & 42.5\% \\ \cline{2-4} 
                        & Male                             & \multicolumn{1}{r|}{37.3\%} & 57.5\% \\ \cline{2-4} 
                        & Other                            & \multicolumn{1}{r|}{1.8\%}  & 0.0\%  \\ \cline{2-4} 
                        & Prefer not to answer                              & \multicolumn{1}{r|}{0.9\%}  & 0.0\%  \\ \hline
\multirow{6}{*}{Age}    & 18$\sim$29                       & \multicolumn{1}{r|}{13.6\%} & 11.3\% \\ \cline{2-4} 
                        & 30$\sim$39                       & \multicolumn{1}{r|}{34.5\%} & 21.7\% \\ \cline{2-4} 
                        & 40$\sim$49                       & \multicolumn{1}{r|}{18.2\%} & 22.6\% \\ \cline{2-4} 
                        & 50$\sim$59                       & \multicolumn{1}{r|}{20.0\%} & 19.8\% \\ \cline{2-4} 
                        & 60 or above                      & \multicolumn{1}{r|}{13.6\%} & 24.5\% \\ \cline{2-4} 
                        & Prefer not to answer                              & \multicolumn{1}{r|}{0.0\%}  & 0.0\%  \\ \hline
\multirow{6}{*}{Education} &
  High school degree or   less &
  \multicolumn{1}{r|}{24.5\%} &
  30.2\% \\ \cline{2-4} 
                        & Associate degree                 & \multicolumn{1}{r|}{17.3\%} & 16.0\% \\ \cline{2-4} 
                        & Bachelor's degree                & \multicolumn{1}{r|}{43.6\%} & 38.7\% \\ \cline{2-4} 
                        & Graduate degree                  & \multicolumn{1}{r|}{11.8\%} & 11.3\% \\ \cline{2-4} 
                        & Others                           & \multicolumn{1}{r|}{2.7\%}  & 3.8\%  \\ \cline{2-4} 
                        & Prefer not to answer                              & \multicolumn{1}{r|}{0.0\%}  & 0.0\%  \\ \hline
\multirow{7}{*}{Ethnicity} &
  White / Caucasian &
  \multicolumn{1}{r|}{72.7\%} &
  87.7\% \\ \cline{2-4} 
                        & Black / African American         & \multicolumn{1}{r|}{15.5\%} & 3.8\%  \\ \cline{2-4} 
                        & Asian                            & \multicolumn{1}{r|}{9.1\%}  & 4.7\%  \\ \cline{2-4} 
                        & Hispanic / Latino                & \multicolumn{1}{r|}{5.5\%}  & 6.6\%  \\ \cline{2-4} 
                        & American Indian / Alaskan Native & \multicolumn{1}{r|}{0.9\%}  & 0.9\%  \\ \cline{2-4} 
                        & Other                            & \multicolumn{1}{r|}{0.9\%}  & 0.0\%  \\ \cline{2-4} 
                        & Prefer not to answer                              & \multicolumn{1}{r|}{0.0\%}  & 0.0\%  \\ \hline
\end{tabular}%
}
\caption{Demographic information of the 216 participants in the main study, categorized by political stance. Participants were allowed to select multiple ethnicities, which result in percentages exceeding 100\% in the ethnicity category.}\label{tab:demo}
\end{table}

\subsection{Participants}
We designed our study using Qualtrics and published it on the online crowdsourcing platform, Prolific, to recruit participants in April 2024. The study involved a set of news items primarily focused on U.S. politics, written in English, and formatted as Facebook posts. Thus, to ensure relevance and quality of responses, we prescreen participants based on several criteria: current residency in the United States, U.S. nationality, English as a first language, over 10 years of residence in the U.S., age over 18, monthly use of social media platforms such as Facebook or X, completion of more than 10 tasks on Prolific, and a minimum approval rate of 95\% or higher. Additionally, each Prolific worker is allowed only one participation in our study. The study took approximately 7 minutes to complete, so we paid each participant \$1 upon completion of our survey. We obtained IRB approval.

To determine the sample size for our main study, we conducted a power analysis using G-Power 3.1 \cite{faul2007g}. Focusing on the accuracy of fake news, we used a 3 (\textit{fact-checker condition}: Congruent vs. Incongruent vs. No Fact-checker) × 2 (News Leaning: Congruent vs. Incongruent) mixed ANOVA. Assuming a median effect size (f= 0.25) for the fact-checking warning effects~\cite{martel2023misinformation}, with an alpha level of 0.05 and a power of 0.80. This analysis suggested a requirement of 120 participants. 
To ensure sufficient power for subgroup analyses and account for variability in online studies conducted on Prolific, we doubled the number and published 240 tasks. Using the prescreening function of Prolific, we recruited 120 participants whose U.S. political affiliation is Democrat and another 120 whose affiliation is Republican. Additionally, participants who took part in the pretest were excluded from the main study to prevent bias.

To maintain data quality, we established exclusion criteria: we excluded two responses due to duplicate IP addresses and two for selecting the same answer across all 12 news items (i.e., straight-lining). Additionally, at the end of our study, we asked participants to self-identify their political stance on a 5-point scale from Very Liberal to Very Conservative. Twenty participants showed contradictions between their political affiliations on Prolific and their responses in our study (e.g., listed as Democrat on Prolific but identified as Conservative in our study). Consequently, we excluded a total of 24 responses, leaving 216 for data analysis. 

\subsection{Procedure}
\begin{figure}[h!]%[h!]
  \centering
  \includegraphics[width=0.475\textwidth]{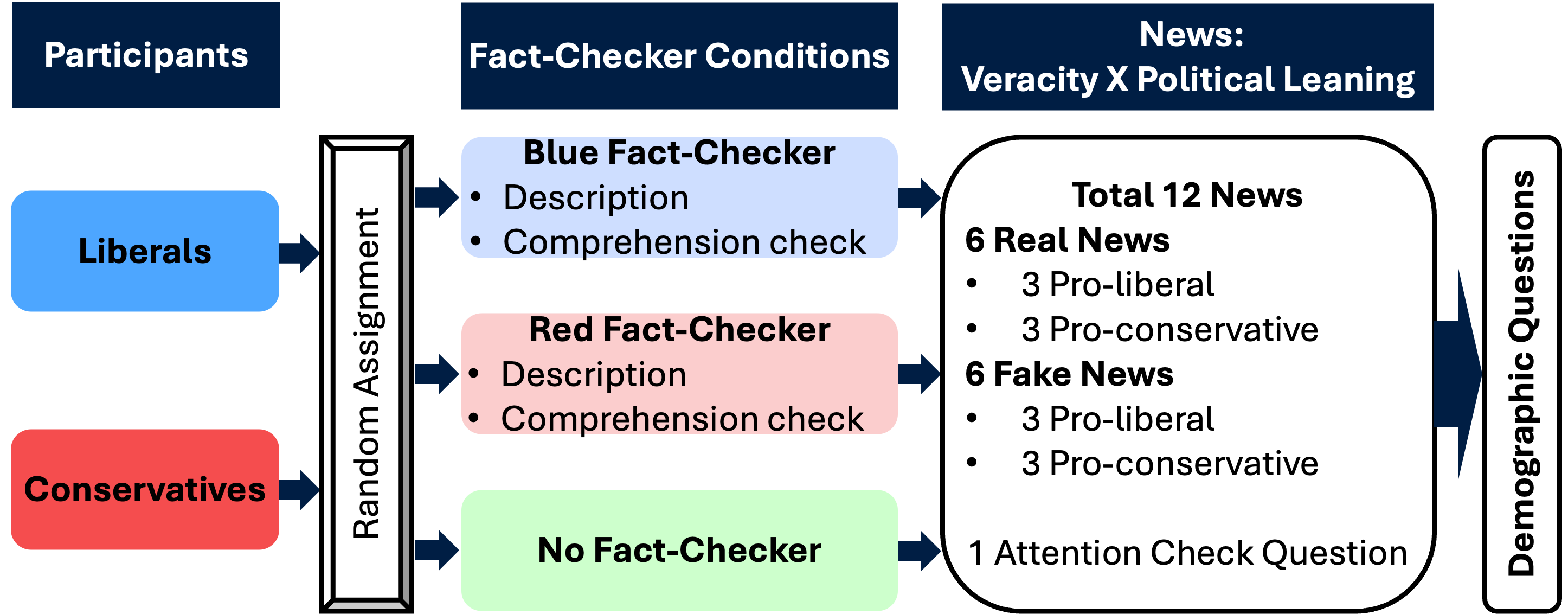}
  \caption{Overview of the Study Flowchart.}
  \label{fig:flow}
\end{figure}

Only participants who met the prescreening criteria (see Participants section) were eligible to participate in our task on Prolific. After accepting the task, they were directed to an online survey hosted on Qualtrics (see Figure~\ref{fig:flow} for the study flowchart). 
Participants first reviewed and provided consent through a consent form. Following this, they were randomly assigned to one of three conditions: \textit{Blue Fact-Checker}, \textit{Red Fact-Checker}, or \textit{No Fact-Checker}. All participants viewed the same 12 news items (6 real, 6 fake), evenly split between pro-liberal and pro-conservative viewpoints, presented as Facebook posts (see Figure~\ref{fig:fbpost}) in random order. Participants rated the accuracy of each headline on a 7-point scale from Very Inaccurate (1) to Very Accurate (7). An attention check was included, requiring participants to select the specified correct option for one randomly presented question in addition to the 12 posts.

\begin{figure}[h!]%[h!]
  \centering
  \includegraphics[width=0.3\textwidth]{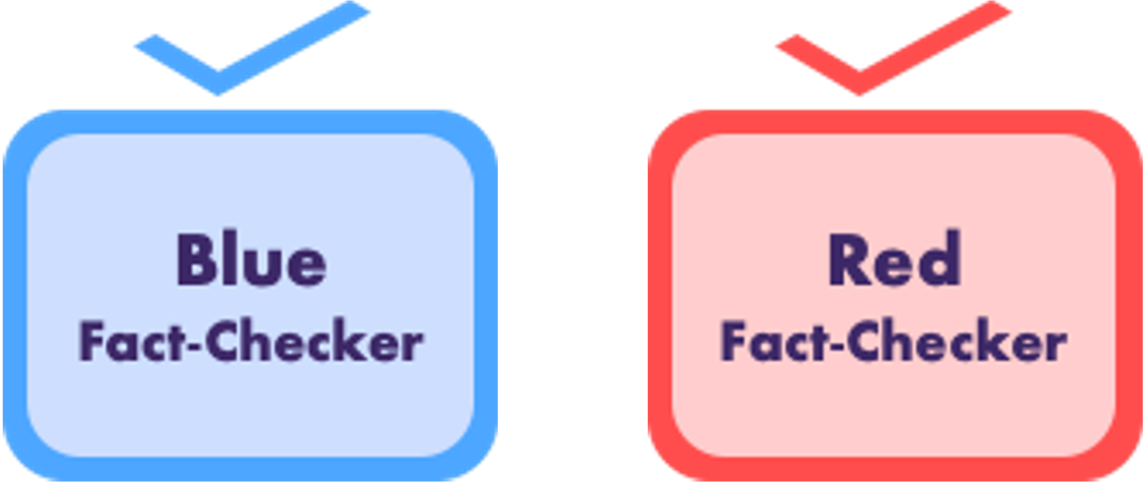}
  \caption{Icons used for the Blue Fact-Checker (left) and the Red Fact-Checker (right).}
  \label{fig:fcmark}
\end{figure}

The presentation of fact-checking messages for fake news varied depending on the condition. In the \textit{No Fact-Checker} condition, no warning tags or messages were displayed, and participants evaluated the posts based solely on their judgment of the headlines. In contrast, the \textit{Blue} and \textit{Red} conditions included a warning tag and a fact-checking message accompanying the fake news posts.
Before viewing the posts, participants saw a description about the biases of their assigned Fact-Checker: those in the \textit{Blue} condition saw an explanation describing the \textit{Blue Fact-Checker} as leaning liberal and often labeling statements from conservative politicians as false, while those in the \textit{Red} condition saw an explanation describing the \textit{Red Fact-Checker} as leaning conservative and frequently labeling statements from liberal politicians as false~\cite{brandenburg2006party, d2000media, eberl2017one, groeling2013media, haselmayer2017partisan, hofstetter1978bias, lichter2017theories}. Example headlines illustrating these biases were also provided (see supplementary material for the full description of each Fact-Checker). In the No Fact-Checker condition, no explanation about fact-checkers was given. These explanations were accompanied by the corresponding Blue or Red Fact-Checker icons (see Figure 3).

After reading the fact-checker description, participants rated the perceived favorability of their assigned Fact-Checker towards Democrats versus Republicans as a comprehension check. For example, participants in the Blue condition were asked: \textit{``Based on the explanation above, how would you rate the Blue Fact-Checker's favorability towards Democrats versus Republicans?''} The rating scale ranged from `Very favorable to Democrats (1),' through `Neutral (4),' to `Very favorable to Republicans (7).' If their response did not align with the provided description (e.g., those in the Blue condition who indicated that the Blue Fact-Checker is neutral or favorable towards Republicans), they were prompted to review the description again before proceeding. Of the 216 participants, 17 (7.9\%) initially misunderstood the partisan alignment of the fact-checkers. Participants whose responses aligned with the description continued directly to the news rating task.

Participants evaluated the accuracy of the 12 headlines, with those in the \textit{Blue} and \textit{Red} conditions seeing additional fact-checking tags and messages for the 6 fake news posts. The content of these messages was consistent across both conditions and based on previously debunked articles from fact-checking websites (see Figure~\ref{fig:fbpost} for an example). Finally, participants completed demographic questions and self-identified their political stance (see Table~\ref{tab:demo}). 

\subsection{Data Analysis}
The dependent measure in our main study is participants' \textit{perceived accuracy ratings} of the headlines. Our study includes four independent measures: 1) \textit{veracity} (real, fake), 2) participants' self-identified \textit{political ideology} (liberal, conservative), 3) \textit{news stance} (pro-liberal, pro-conservative), and 4) \textit{fact-checker condition} (blue, red, no fact-checker). 

We first examined how political stance congruency impacts participants' perceived accuracy ratings. We calculated the average perceived accuracy ratings based on participants' self-identified \textit{political ideology} aligned with \textit{news stance} (i.e., congruent news vs. incongruent news), and \textit{fact-checker condition} (i.e., congruent fact-checker vs. incongruent fact-checker vs. no fact-checker). For example, for conservative participants, the average perceived accuracy rating of pro-conservative news was calculated as congruent news, while that of pro-liberal news was calculated as incongruent news, and vice versa for liberal participants. Similarly, for conservative participants assigned to the Red Fact-Checker condition, this was considered the congruent fact-checker condition, those in the Blue Fact-Checker condition were considered to be in the incongruent fact-checker condition, and those in the no fact-checker condition remained as such.

Later, to examine differences across various political stances, we dissected the congruency into each specific political stance and included all four independent measures in our analysis of the effects of political stance.

\section{Result}
\subsection{Analysis of Congruency Effects}
We conducted a mixed-design ANOVA with a 2 (\textit{veracity}: real, fake) $\times$ 2 (\textit{news stance}: congruent news, incongruent news) $\times$ 3 (\textit{fact-checker condition}: congruent fact-checker, incongruent fact-checker, no fact-checker) factorial structure, analyzing perceived accuracy ratings at a significance level of 0.05. Post-hoc analyses with Bonferroni corrections were performed to control for potential inflation of p-values in pairwise comparisons.

\begin{figure*}[t!]%[h!]
  \centering
  \includegraphics[width=1\textwidth]{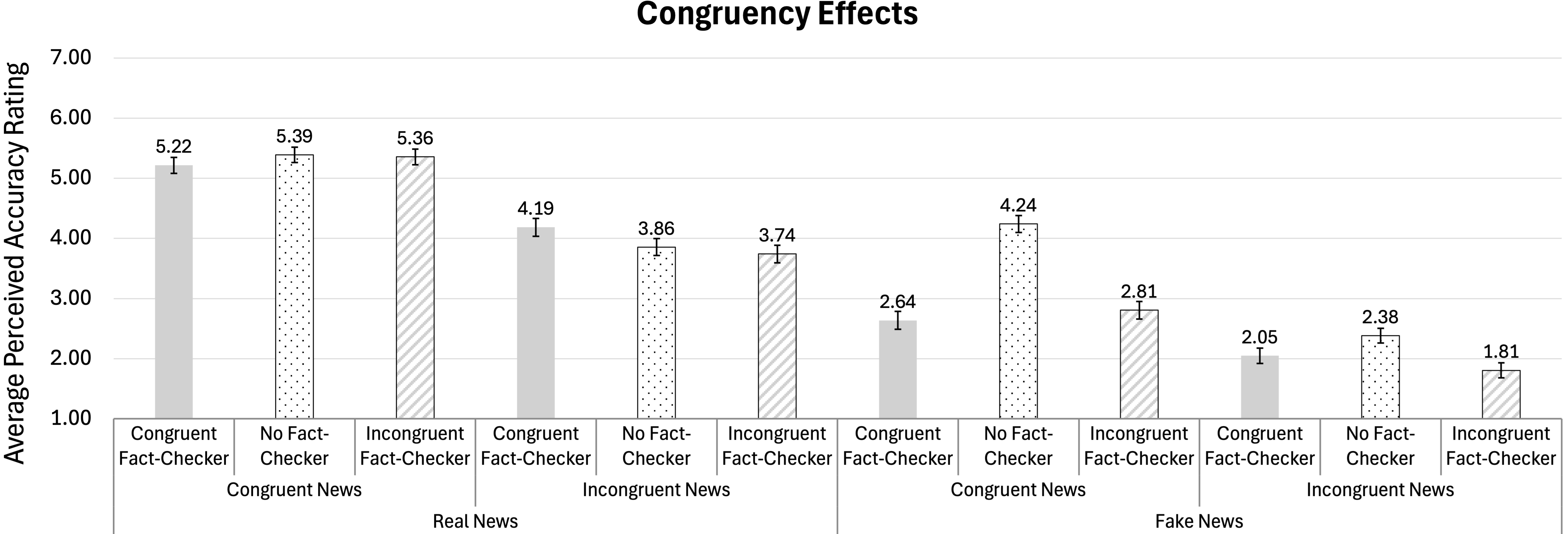}
  \caption{Average accuracy ratings across a 2 (\textit{veracity}: real, fake) $\times$ 2 (\textit{news stance}: congruent, incongruent) $\times$ 3 (\textit{fact-checker condition}: congruent fact-checker, incongruent fact-checker, no fact-checker) factorial design in the congruency analysis. Error bars represent $\pm$ one standard error. The analysis included 68 participants in the \textit{congruent fact-checker}, 72 in the \textit{incongruent fact-checker}, and 76 in the \textit{no fact-checker} condition.}
  \label{fig:cong}
\end{figure*}

Figure~\ref{fig:cong} depicts the results. The main effect of veracity was significant, $F(1, 213) = 589.87, p < .001, \textit{$\eta_{p}^{2}$} = .735$, showing that participants can separate real (4.62) from fake (2.65) news clearly. Also, the main effect of  \textit{fact-checker condition} was significant, $F(2, 213) = 16.58, p < .001, \textit{$\eta_{p}^{2}$} = 0.135$. Pairwise comparison results showed that regardless of whether it is congruent (3.52) or incongruent (3.43) fact-checker ($p_{adj.} = 1.00)$, participants gave lower accuracy ratings compared to no fact-checker condition (3.97) in general ($p_{adjs.} < .001$). 
The two-way interaction between \textit{veracity} and \textit{fact-checker condition} was significant, $F(2, 213) = 17.08, p < .001, \textit{$\eta_{p}^{2}$} = 0.138$. For real news, no significant differences were observed among the fact-checker conditions (congruent, incongruent, and no fact-checker; $p_{adjs.} > .834$).
 
However, fact-checker's warning messages significantly lower the participants' perceived accuracy rating of  fake news, whether it is a congruent (2.34) or incongruent (2.31) fact-checker, compared to no fact-checker condition (3.31, $p_{adjs.} < .001$). 
These results demonstrate a significant effect of fact-checking warnings even with clear political labels on the fact-checkers, regardless of whether they are congruent or incongruent with the participants' political stances.

The main effect of \textit{news stance} was significant, $F(1, 213) = 202.07, p < .001, \textit{$\eta_{p}^{2}$} = .487$, showing participants gave significantly higher accuracy ratings for the congruent news (4.27) compared to incongruent news (3.00). This participants' congruency effect towards news stance was more evident for real news compared to fake news, $F(1, 213) = 4.43, p = .036, \eta_p^2 = 0.020$. Specifically, the difference in perceived accuracy ratings between congruent and incongruent news was larger for real news (5.32 vs. 3.93) than for fake news (3.23 vs. 2.08).

The two-way interaction between \textit{news stance} and \textit{fact-checker condition} was significant, $F(2, 213) = 8.23, p < .001, \eta_p^2 = .072$. For congruent news, the presence of a fact-checker, whether congruent (3.93) or incongruent (4.08), significantly reduced participants' perceived accuracy ratings compared to the no fact-checker condition (4.82, $p_{adjs.} < .001$). However, for incongruent news, only the incongruent fact-checker (2.77) significantly reduced perceived accuracy ratings compared to the no fact-checker condition (3.12, $p_{adj} = .046$), and was marginally lower than the congruent fact-checker condition (3.12, $p_{adj} = .056$). There was no significant difference between the congruent and no fact-checker conditions ($p_{adj} = 1.00$).

To further explore these interactions, we examined the three-way interaction of \textit{veracity} × \textit{news stance} × \textit{fact-checker condition}, which was significant, $F(2, 213) = 6.40, p = .002, \eta_p^2 = .057$. Post-hoc analysis showed that the interaction between \textit{news stance} and \textit{fact-checker condition} approached significance for real news, $F(2, 213) = 2.62, p = .075, \eta_p^2 = .024$, but was significant for fake news, $F(2, 213) = 13.84, p < .001, \eta_p^2 = .115$. Pairwise comparisons found no significant differences in accuracy ratings for real news across fact-checker conditions, either for congruent news (congruent fact-checker vs. incongruent fact-checker vs. no fact-checker: 5.22 vs. 5.36 vs. 5.39, $p_{adjs.} = 1.00$) or for incongruent news (4.19 vs. 3.74 vs. 3.86, $p_{adjs.} \geq 0.099$). However, for fake news, fact-checker warnings significantly affected accuracy ratings. For congruent fake news, both congruent (2.64) and incongruent (2.81) fact-checkers significantly lowered accuracy ratings compared to no fact-checker (4.24, $p_{adjs.} < .001$), with no significant difference between the types of fact-checkers ($p_{adj.} = 1.00$). For incongruent fake news, only the incongruent fact-checker (1.81) significantly reduced accuracy ratings compared to no fact-checker (2.38, $p_{adj.} = .003$), and the congruent fact-checker (2.05) did not differ significantly from other conditions ($p_{adjs.} \geq .184$). 

This suggests that for congruent fake news, participants accepted the fact-checker’s warning messages regardless of the fact-checker’s congruence, adjusting their higher belief in the congruent fake news downward, even when they perceived the fact-checker as biased. In contrast, participants were already critical of the incongruent fake news and also perceived the congruent fact-checkers as biased. As a result, they may have viewed the congruent fact-checker as excessively critical of the incongruent news compared to the incongruent fact-checker. This perception could lead to lower acceptance of the fact-checking message from the congruent fact-checker, making it indistinguishable from the no fact-checker condition. Conversely, the incongruent fact-checker, possibly perceived as an `unlikely' source and thus more objective, resulted in greater acceptance of the fact-checking message, showing a significant difference from the no fact-checker condition.

\begin{figure}[h!]%[h!]
  \centering
  \includegraphics[width=\columnwidth]{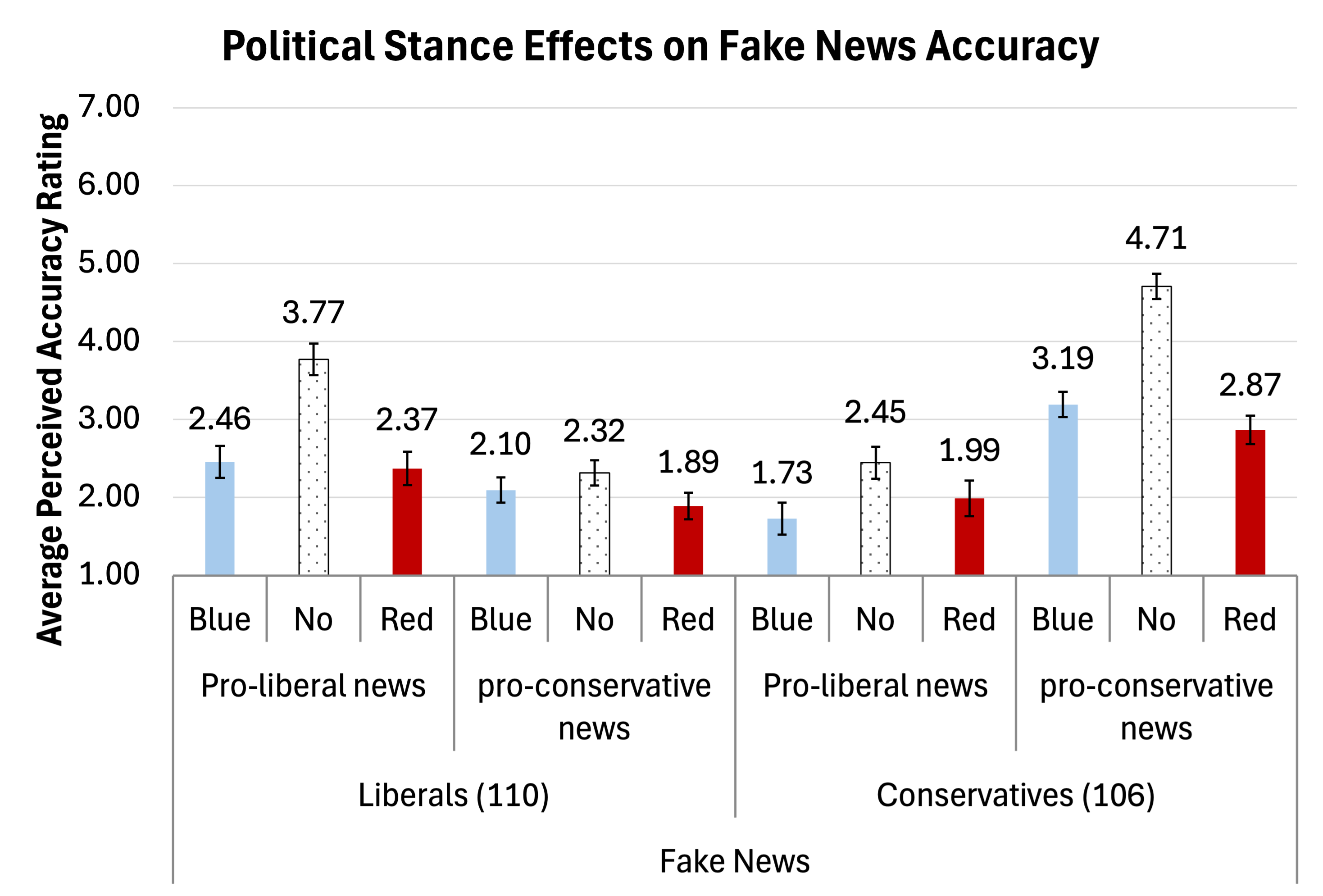}
  \caption{Average accuracy ratings of fake news across a 2 (\textit{veracity}: real, fake) × 2 (\textit{political ideology}: liberal, conservative) × 2 (\textit{news stance}: pro-liberal, pro-conservative) × 3 (\textit{fact-checker condition}: Blue, Red, no fact-checker) factorial design in the political stance analysis. Error bars represent $\pm$ one standard error. Among liberals, there were 38 participants in each of the \textit{Blue} and \textit{no fact-checker} conditions, and 34 in the \textit{Red} fact-checker condition. Among conservatives, there were 38 participants in each of the \textit{Blue} and \textit{no fact-checker} conditions, and 30 in the \textit{Red} fact-checker condition.}
  \label{fig:poli}
\end{figure}

\subsection{Analysis of the Effects of Political Stance}
To further understand the effect of political stance on the impact of fact-checking warnings, we examined the congruence in terms of its original political orientations (i.e., liberal or conservative). For this analysis, we conducted 2 (\textit{veracity}: real, fake) × 2 (\textit{political ideology}: liberal, conservative) × 2 (\textit{news stance}: pro-liberal, pro-conservative) × 3 (\textit{fact-checker condition}: Blue, Red, no fact-checker) mixed ANOVA, with perceived accuracy ratings as the dependent variable. Here, we report analysis results that provide further findings beyond our previous analysis.

The two-way interaction of \textit{veracity} and \textit{political ideology} was significant, $F(1, 210) = 13.67, p < .001, \eta_p^{2} = .061$, highlighting that liberal participants gave higher accuracy ratings for real news (liberals vs. conservatives: 4.75 vs. 4.50, $p_{adj.} = .030$) but lower accuracy ratings for fake news (liberal vs. conservative: 2.48 vs. 2.82, $p_{adj.} = .004$) compared to conservative participants. This implies that liberal participants differentiated more clearly between real and fake news compared to conservative participants, consistent with previous research~\cite{garrett2021conservatives,jost2018ideological,swire2017processing}.

Furthermore, the three-way interaction of \textit{veracity} $\times$ \textit{news stance} $\times$ \textit{political ideology} was significant, $F(1, 210) = 4.81, p = .029, \eta_{p}^{2} = .022$. Post-hoc analyses revealed that the two-way interaction was significant only for conservatives, $F(1, 103) = 4.76, p = .031, \eta_{p}^{2} = .044$, but not for liberals, $F<1$. Specifically, the gap in accuracy ratings (i.e., congruency effects of news stance) between pro-liberal and pro-conservative news for liberal participants was similar across both real (5.20 vs. 4.30; gap = .90) and fake news (2.87 vs. 2.10; gap = 0.77). Conversely, conservative participants exhibited larger congruency effects of news stance for real news (pro-liberal vs. pro-conservative: 3.55 vs. 5.46; gap = 1.91) than for fake news (pro-liberal vs. pro-conservative: 2.06 vs. 3.59; gap = 1.53).

The main effect of \textit{news stance} was significant, $F(1, 210) = 27.08, p < .001, \eta_{p}^{2} = 0.114$, with pro-conservative news (3.86) rated higher than pro-liberal news (3.42). The two-way interaction between \textit{news stance} and \textit{political ideology} was also significant, $F(1, 210) = 226.54, p < .001, \eta_{p}^{2} = .519$. This interaction showed that liberals rated pro-liberal news (4.04) higher than pro-conservative news (3.20, $p_{adj.} < .001$), while conservatives showed the opposite pattern (pro-liberal vs. pro-conservative: 2.80 vs. 4.52, $p_{adj.} < .001$).

Furthermore, the three-way interaction of \textit{news stance} $\times$ \textit{fact-checker condition} $\times$ \textit{political ideology} was significant, $F(2, 210) = 6.29, p = .002, \eta_{p}^{2} = .057$, as was the four-way interaction involving \textit{veracity}, $F(2, 210) = 6.87, p = .001, \eta_{p}^{2} = .061$ (see Figure~\ref{fig:poli}). Post-hoc analyses showed that the three-way interaction was significant only for fake news, $F(2, 210) = 13.58, p < .001, \eta_{p}^{2} = .115$, and not for real news, $F<1$, suggesting that fact-checker warnings did not significantly change participants' accuracy ratings of real news for either liberals or conservatives in pro-liberal or pro-conservative contexts ($p_{adjs.}\geq .285$).

However, post-hoc analysis for fake news (Figure~\ref{fig:poli}) revealed that the two-way interaction of \textit{news stance} $\times$ \textit{fact-checker condition} was significant for both liberals, $F(2, 107) = 9.15, p < .001, \eta_{p}^{2} = 0.146$, and conservatives, $F(2, 103) = 6.11, p = .003, \eta_{p}^{2} = 0.106$, albeit with different patterns. Pairwise comparisons showed that for liberals, both Blue (2.46) and Red (2.37) fact-checkers effectively decreased the accuracy ratings of pro-liberal fake news compared to no fact-checker (3.77, $p_{adjs.} < .001$), but not for pro-conservative news (Blue vs. Red vs. no fact-checker: 2.10 vs. 1.89 vs. 2.32, $p_{adjs.} \geq .372$). Conversely, for conservatives, both Blue (3.19) and Red (2.87) fact-checkers significantly lowered the perceived accuracy rating of pro-conservative fake news compared to no fact-checker (4.71, $p_{adjs.} < .001$). This indicates that for both liberal and conservative participants, the fact-checking warning effectively mitigates the congruency effects in congruent fake news.

Additionally, only for conservative participants, the fact-checker condition significantly affected the ratings for pro-liberal fake news, $F(2, 103) = 4.11, p = .019, \eta_{p}^{2} = 0.074$, with the Blue fact-checker further significantly lowering the perceived accuracy rating of pro-liberal fake news (1.73) compared to no fact-checker (2.45, $p_{adj.} = .013$). This suggests that when the fake news is incongruent with the participant's political ideology, presenting a fact-checking warning message from a fact-checker aligned with the political stance of the fake news could be more effective in correcting misinformation, especially for conservatives. We further discuss the possible reason why this occurs only among conservatives in the General Discussion section.

Furthermore, the results showed that without the fact-checker, conservative participants rated the pro-conservative news with high perceived accuracy (4.71), suggesting they considered this fake news to be credible, with ratings above the neutral midpoint of 4. In contrast, although liberal participants also exhibited congruence effects toward pro-liberal news under the no fact-checker condition (3.77), their ratings remained below 4, indicating a tendency to view it as likely fake. Moreover, under the Blue fact-checker condition, conservative participants gave the lowest perceived accuracy rating (1.73) to pro-liberal news, the lowest across all news types and conditions.
\section{General Discussion}
Our study demonstrated that partisan fact-checkers can effectively reduce false beliefs about political misinformation via social media fact-checking messages, regardless of whether their political views congruent or incongruent with the users' political ideologies (\textbf{RQ1}), without triggering any backfire effects. Additionally, our results indicated that fact-checking interventions are particularly effective at addressing misinformation that aligns with the participants' own political views compared to those that do not (\textbf{RQ2}). When analyzing the influence of political stance, we noted that the 
Blue fact-checker, biased toward liberal views, was more effective at mitigating conservative individuals' misbeliefs about pro-liberal misinformation.
Such an effect, however, was not evident among liberal participants (\textbf{RQ3}).

\subsection{Partisan Fact-Checkers Correct Political Misinformation Without Inducing Backfire}

Throughout our experiment, we showed that partisan fact-checkers, whether biased toward liberal or conservative views, can reduce people's perceived accuracy ratings of political misinformation and correct their misbeliefs. 
Even in politically polarized contexts, where individuals strongly align with misinformation, fact-checkers remained effective despite perceptions of bias in the media, including among fact-checkers. Moreover, while political misinformation tends to be especially persistent~\cite{bode2015related,garrett2013undermining,jerit2012partisan,taber2006motivated}, our findings showed that partisan fact-checkers remain effective in combating misinformation through the presentation of fact-checking warning messages on social media posts that contain political misinformation.

Furthermore, while some studies have shown that fact-checking warning labels may be less effective for conservatives than for liberals~\cite{jennings2023asymmetric,morris2020fake,yaqub2020effects}, our results suggest that such messages on social media can mitigate misbeliefs about political misinformation for both groups, even when the source of the warning (i.e., the fact-checker) is perceived as politically biased. This finding is consistent with previous research showing that fact-checking warnings are generally effective regardless of political alignment~\cite{martel2023misinformation, porter2022political, swire2017processing, swire2020they}. Our results build on this foundation by demonstrating that politically biased fact-checkers can still reduce the perceived accuracy of political misinformation.

While \citeauthor{nyhan2010corrections} (\citeyear{nyhan2010corrections}) reported possible `backfire' effects when correcting political misinformation, subsequent research has consistently shown that such backfire effects are unlikely to occur, even when addressing highly polarized issues \cite{ecker2020can, prike2023examining, wood2019elusive}. Our study also confirms that no backfire effects were triggered when correcting political misinformation among polarized participants, including both liberals and conservatives, even when the fact-checking messages originated from politically biased fact-checkers.

\subsection{Fact-Checking Messages Effectively Correct Politically Congruent Misinformation}

Importantly, our analysis comparing correction effects on both pro-liberal and pro-conservative news reveals that fact-checking messages are more effective at correcting misinformation that aligns with people's political ideologies than misinformation that misaligns with them, for both conservatives and liberals. This finding highlights the effectiveness of fact-checking messages on social media in mitigating the congruency effects associated with politically aligned misinformation. Given the increased susceptibility to politically congruent misinformation, which may stem from a `laziness' in reasoning rather than motivated reasoning~\cite{pennycook2019lazy}, participants likely employed more reasoning for incongruent headlines and resorted to heuristics for congruent but implausible (i.e., fake) headlines. Therefore, the increased susceptibility to congruent fake news could be attributed to a lack of reasoning rather than motivated reasoning. Consequently, fact-checking messages could promote reasoning-based accuracy rating decisions for congruent fake news, correcting it more effectively.

\subsection{Partisan Counterparts Further Reduce Misbeliefs Among Conservatives: A Novel Finding More Effective Than for Liberals}
Another intriguing result from our study is that for conservatives, only the Blue fact-checker significantly reduced the perceived accuracy ratings of pro-liberal misinformation. This aligns with previous research which suggests that corrections from `unlikely sources' (e.g., a Republican debunking statements from another Republican) can enhance individuals' willingness to dismiss such rumors~\cite{berinsky2017rumors,calvert1985value,petty1986elaboration}. 
This result highlights the role of the messenger effect, where the perceived identity and alignment of the fact-checker influence the reception of their message~\cite{mcginnies1980better,petty1986elaboration}. The effectiveness of the Blue fact-checker in addressing pro-liberal misinformation among conservatives likely arises from the `unexpectedness' of a fact-checker perceived as liberal correcting information that aligns with liberal ideologies. This unexpected action may challenge pre-existing biases and encourage greater receptiveness to the correction.
\begin{figure}[h!]%[h!]
  \centering
  \includegraphics[width=0.47\textwidth]{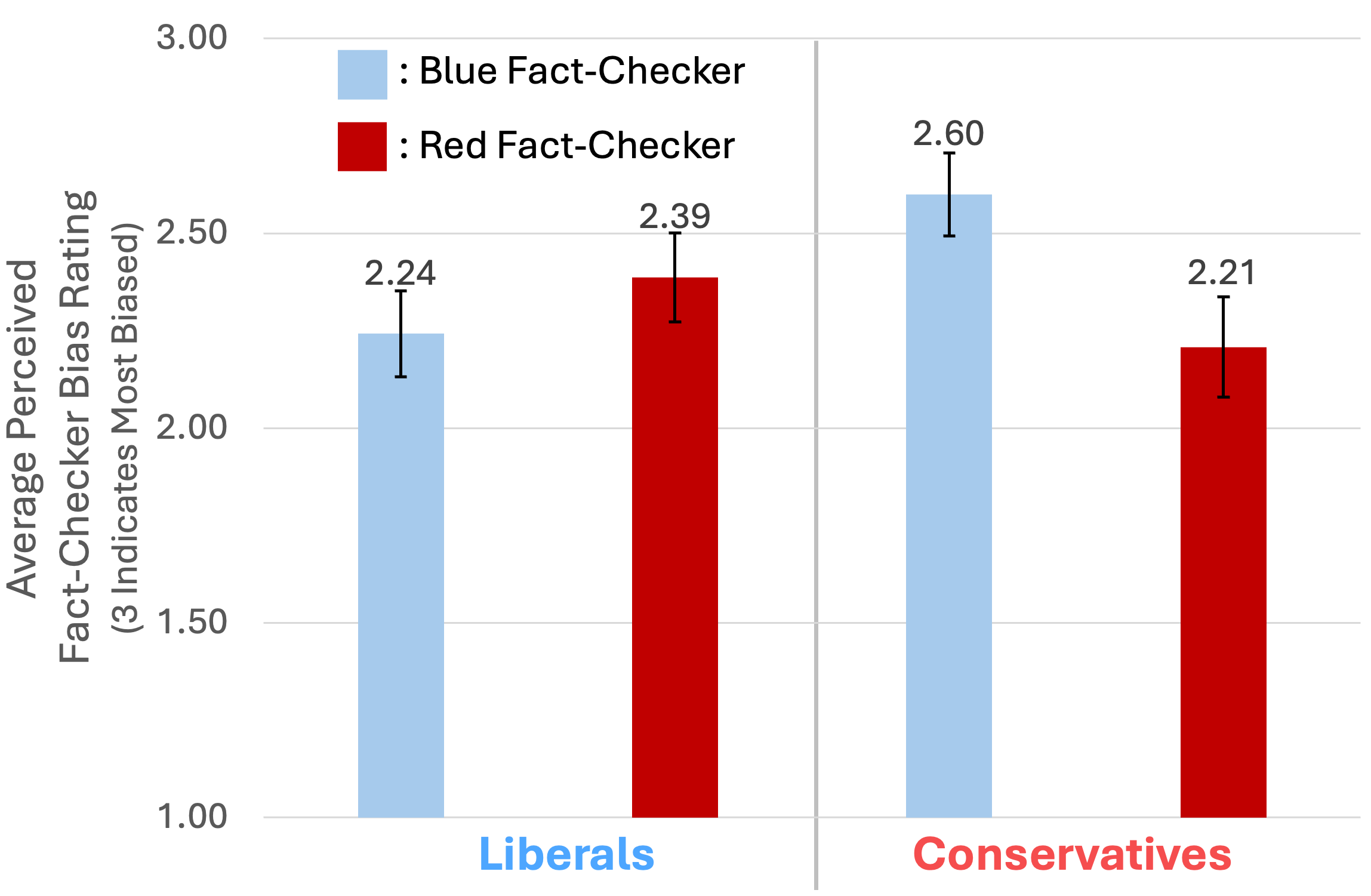}
  \caption{Fact-Checker Bias Rating. Higher ratings indicate that participants perceive the fact-checker as more biased.}
  \label{fig:fcbias}
\end{figure}

Although a numerically similar pattern was observed for liberals, specifically when the Red fact-checker debunked pro-conservative misinformation (see Figure 5), this effect did not reach statistical significance among liberal participants. A possible explanation is that conservative participants are more likely to perceive the Blue fact-checker as biased towards liberal views compared to how liberal participants perceive the Red fact-checker as biased towards conservative views (see Figure~\ref{fig:fcbias}). Our data also show that liberal participants assigned to the Blue fact-checker condition rated the fact-checker's favorability at 1.76 on a scale from 1 to 7, where 1 represents a bias towards liberal views and 7 towards conservative views (for more details, see the Procedure section). Participants in the Red fact-checker condition rated it at 6.39. These ratings illustrate the liberals' perceived political bias for both the Blue and Red fact-checkers, with deviations from the neutral midpoint of 4 being 2.24 and 2.39, respectively, as shown in Figure~\ref{fig:fcbias}. Conversely, conservative participants assigned to the Blue fact-checker condition perceived a pronounced bias, rating it at 1.40. This reflects a bias rating of 2.60 from the neutral midpoint of 4, indicating the highest perceived level of bias across all conditions. Meanwhile, the Red fact-checker was seen as less biased, with a bias strength of 2.21 (rating 6.21). This suggests that conservatives perceive the Blue fact-checker's debunking of pro-liberal news as highly `unlikely,' thereby increasing their receptiveness to corrections issued by the Blue fact-checker. However, given the small sample size, caution is advised in interpreting these results.

Fact-checking efforts in the U.S. are often criticized for a perceived `left-leaning' bias, which may arise from asymmetries in misinformation sharing across political groups~\cite{mosleh2024differences}. Our findings emphasize the importance of addressing these perceptions to ensure the effective implementation of fact-checking interventions. The results highlight the significant role of messenger effects, demonstrating how perceived partisan alignment influences the acceptance of corrections. Notably, the `unexpectedness' of corrections from partisan counterparts may serve as a potentially effective mechanism for reducing misbeliefs, particularly in politically polarized contexts.

\subsection{Limitations and Future Directions}
Although we aimed to enhance ecological validity, numerous real-world factors—including the number of likes, shares, and the identity of the poster—can significantly influence the acceptance of corrections for misinformation. Additionally, \citeauthor{margolin2018political} (\citeyear{margolin2018political}) demonstrated that reciprocal relationships between fact-checkers and recipients can substantially improve the acceptance of corrections—a dynamic not addressed in our non-interactive experimental setting. \citeauthor{parekh2020comparing} (\citeyear{parekh2020comparing}) further demonstrated that the reception and impact of fact-checking vary across online communities, with corrections often being more appreciated in less partisan environments. To gain a deeper understanding of the effectiveness of corrections, future studies should leverage real-world social media data to better capture the influence of these factors.

Another limitation concerns demographic differences between liberals and conservatives. While our study's demographics somewhat mirror those of U.S. liberals and conservatives — with liberals generally being younger, more educated, more often female, and having a lower proportion of white individuals compared to conservatives ~\cite{pew2024demo} — the relatively small sample size constrained our ability to fully account for the potential impact of these demographic factors. Future studies should aim to increase the sample size and incorporate more robust controls for demographic variables to better understand their influence on responses to fact-checking interventions.

Furthermore, we intentionally selected headlines published at least 10 months prior to the study to minimize the likelihood of participants being familiar with them, control other variables, and specifically highlight the effects of partisan fact-checking within a controlled experimental setting. However, we acknowledge that this approach may not fully capture reactions to current news events. To address this limitation, future analyses incorporating real-world, up-to-date headlines are recommended to enhance ecological validity.

Another limitation of our study is its design with a focus on immediate responses to fact-checking interventions without examining how these corrections influence beliefs over time. Prior research highlights that longitudinal approaches are crucial for understanding the durability of fact-checking effects~\cite{rich2020correcting}. Future studies could explore whether the effects of corrections from partisan fact-checkers persist in the long term or diminish over time. Moreover, repeated corrections from the same or different partisan fact-checkers could offer valuable insights into cumulative effects and potential diminishing returns in belief updating \cite{prike2023effective}. Investigating these factors would contribute to a more comprehensive understanding of how fact-checking interventions function in complex, real-world settings.
\section{Conclusion}
Our study demonstrates that partisan fact-checkers can reduce people's perceived accuracy of political misinformation and correct misbeliefs through fact-checking warnings presented on social media, all without triggering backfire effects. Moreover, this effect was even more pronounced when correcting misbeliefs about misinformation that aligns with individuals' political ideologies. Importantly, contrary to the notion that fact-checking warnings are less effective for conservatives than liberals, our findings suggest that explicitly labeled partisan fact-checkers, which act as political counterparts to conservatives, could further reduce conservatives' misbeliefs towards pro-liberal misinformation.

\section{Acknowledgements}
This research was supported in part by the National Science Foundation under grants 1820609, 1915801, and 2121097.

\bibliography{aaai25}

\end{document}